\documentclass[final,5p,times,twocolumn]{elsarticle}
\usepackage{graphicx}
\usepackage{amssymb}
\journal{Physica E}

\begin{document}

\begin{frontmatter}
\title{Interaction and temperature effects on the pair correlation function of a strongly interacting 1D quantum dot}
\author[difi,spin]{N. Traverso Ziani}
\author[difi,spin]{F. Cavaliere}
\author[exeter]{E. Mariani}
\author[difi,spin]{M. Sassetti}
\address[difi]{Dipartimento di Fisica, Universit\'a di Genova, Via Dodecaneso 33, 16146 Genova (Italy)}
\address[spin]{CNR-SPIN, Via Dodecaneso 33, 16146 Genova (Italy)}
\address[exeter]{Centre for Graphene Science, School of Physics, University of Exeter, Stocker Rd., EX4 4QL Exeter (UK)}
\begin{abstract}
The pair correlation function of a strongly interacting one-dimensional quantum dot is evaluated analytically within the framework of the spin coherent Luttinger liquid model. The influence of electron interactions and temperature on the competition between finite-size effects and electronic correlations is analyzed, also in the regime of large numbers of particles. The development of Wigner molecule correlations is observed as interactions get stronger. The visibility of such signatures is enhanced for temperatures comparable with the spin excitation energy of the system due to a suppression of the uncorrelated Friedel contribution to the pair correlation function.
\end{abstract}
\end{frontmatter}

\section{Introduction}
\label{sec:intro}
After more than 75 years of investigations, both the theoretical description and the experimental realization of Wigner crystals and Wigner molecules, a paradigmatic example of strongly correlated state of electrons in condensed matter, still represent a challenging problem~\cite{wigner}.\\

\noindent Much insight in this fascinating problem has been gained thanks to quantum dots~\cite{koudots}, an ideal playground for the study of correlated electron systems. In circular two dimensional (2D) quantum dots Wigner molecules have been extensively studied theoretically~\cite{wigmol1,wigmol2}. Here, the electron density only shows weak signatures of the Wigner molecule through a radial ordering, while no angular correlations can be detected due to the rotational symmetry of the system. The latter can only be probed by means of density-density correlation functions or pair correlation functions (PCF). Several different theoretical tools are employed to study 2D quantum dots but all rely heavily on numerical calculations~\cite{2Dn1,2Dn2,2Dn3,2Dn4,2Dn5,2Dn6,2Dn7,2Dn8,2Dn9,2Dn10,2Dn11}, which limits the most accurate investigations to a rather low number of particles $N\sim 10$. Experimentally, the key tool to detect Wigner molecules in 2D dots is optical-spectroscopy, with results being compared with the numerical predictions of the theory~\cite{wigspec}. Also the coupling to local probes such as AFM or STM tips has been proposed~\cite{maxwf,eroszba}.\\

\noindent One-dimensional (1D) systems are qualitatively different, in that breaking the translational invariance makes it possible to deserve strong oscillations already at the level of the electron density. Finite-size effects show up as Friedel oscillations with a wavevector $k_{F}$ (the Fermi momentum of the dot), while electron correlations induce Wigner oscillations characterized by twice the wavevector, i.e. $2k_{F}$~\cite{vignale,1Dwig}. Even though 1D systems have also been approached by means of numerical techniques~\cite{1Dn1,1Dn2,1Dn3,1Dn4,1Dn5,1Dn6,1Dn7,1Dn8,bortz,1Dn9,1Dn10,1Dn11,1Dn12} the peculiar nature of the low-energy excitations of interacting 1D systems, described by the Luttinger liquid theory~\cite{giamarchi}, allows to employ bosonization techniques to obtain analytical results~\cite{lu1,lu2,lu3,lu4} also concerning the transport properties of such systems~\cite{lu5,lu6,lu7,lu8,lu9,lu10,lu11}. The analytical study of the electron density of an interacting 1D quantum dot within the Luttinger liquid theory is particularly simple at zero temperature: the Friedel and the Wigner contributions are well separated and their relative importance is governed by the interaction parameter of the theory~\cite{lu8,lu9}. For increasing temperatures, the thermal activation of spin excited states leads to a suppression of the Friedel term and the emergence of clearer Wigner signatures~\cite{noiNJP13}. It is however interesting to assess whether this behaviour is observed also in two-particle probes such as the PCF, which are a direct measure of the degree of correlation induced by electronic interactions. Indeed, ab initio calculations confirm this quantity as a sensitive tool to assess the degree of correlations also in 1D systems~\cite{lu1,wang}.\\

\noindent In this work we analytically evaluate the PCF of a strongly interacting 1D quantum dot, in the presence of both Friedel and Wigner oscillations. Employing the spin-coherent Luttinger liquid picture~\cite{lu3} we include thermal effects up to a temperature $T\lesssim N E_{\sigma}/k_{B}$ with $E_{\sigma}$ the spin excitation energy. Our analytical approach circumvents the typical restriction to low $N$ of purely numerical approaches and allows in principle to easily investigate systems with an arbitrarily large number of particles. Both the Friedel and the Wigner oscillations induce peculiar patterns in the PCF. The Wigner term induces $N-1$ distinct peaks of the PCF, while the Friedel term tends to favor their grouping in pairs. A crossover towards the Wigner regime is observed as interactions are stronger. We find that the PCF is indeed sensitive  to thermal effects: indeed, the grouping induced by the Friedel oscillations tends to disappear leading to an emergence of the Wigner correlations akin to the phenomenon shown by the single-particle density.\\

\noindent The outline of the paper is as follows. In Sec.~\ref{sec:model} we briefly introduce the spin-coherent Luttinger model for the 1D strongly interacting quantum dot, and give analytic expression for the electron density and the PCF. Section~\ref{sec:res} is devoted to the discussion of the results concerning the PCF, while Sec.~\ref{sec:conc} contains our conclusions.
\section{Model and methods}
\label{sec:model}
\subsection{System hamiltonian}
\label{sec:ham}
 Our model consists of a 1D quantum dot of length $L$ with $N_{\rho}$ extra electrons, described within the coherent Luttinger liquid picture~\cite{giamarchi} valid in the low energy sector provided that the temperature satisfies $k_BT< D_\sigma,D_\rho$, with $D_{\rho/\sigma}=N_\rho E_{\rho/\sigma}$ the band width of the spin and charge sectors respectively. Here, $N_{\rho}=N+\Delta N_{\rho}$ where $N$ represents the reference number of electrons in the systems which we will assume even for definiteness. For higher temperature $D_{\sigma}\ll k_{B}T\ll D_{\rho}$ the {\em spin-incoherent} Luttinger liquid theory must be employed~\cite{lu3,noiNJP13}. The hamiltonian is (henceforth, $\hbar=1$)
\begin{equation}
H_{d}=\frac{E_{\rho}}{2}\Delta N_{\rho}^2+\frac{E_{\sigma}}{2}\Delta N_{\sigma}^2+\sum_{\nu=\rho,\sigma}\sum_{n_{q}>0}\varepsilon_{\nu}n_{q}d^\dag_{\nu,n_{q}}d_{\nu,n_{q}}\, .\label{eq:ham}
\end{equation}
The first two terms of Eq.~(\ref{eq:ham}) represent the zero-modes sector of the theory with $E_{\nu}=\pi v_{\nu}/2L g_{\nu}^2$ the charge ($\nu=\rho$) and spin ($\nu=\sigma$) addition energies of the modes $\Delta N_{\rho}=N_{+}+N_{-}-N$ and $\Delta N_{\sigma}=N_{+}-N_{-}$ with $N_{s}$ the total number of eletrons with spin $s$. One has $|\Delta N_{\sigma}|\leq N+\Delta N_{\rho}$. We deal with an isolated quantum dot which exchanges no particles with the environment. As a consequence, the ground state of the dot has $\Delta N_{\rho}=0$ which implies $N_{+}+N_{-}=N$. Furthermore, $v_{\nu}$ is the velocity of the mode $\nu$ and $g_{\nu}$ models the strength of electron interactions. One has $g_{\rho}=g<1$ for repulsive interactions ($g=1$ is the non-interacting limit) while $g_{\sigma}=1$ is implied by SU(2) invariance. For strong electron interactions one has $v_{\sigma}\ll v_{\rho}$, while $v_{\sigma}=v_{\rho}=v_{F}$ in the non-interacting regime, with $v_{F}$ the Fermi velocity of the system. The strong separation of the spin and charge propagation velocities is responsible for the temperature-induced emergence of Wigner correlations over finite-size effects that we are going go discuss in this paper~\cite{noiNJP13}.\\
\noindent The last term in Eq.~(\ref{eq:ham}) describes collective charge and spin density waves with quantized momentum $q=\pi n_{q}/L$ where $n_{q}>0$ is an integer and $\varepsilon_{\nu}=\pi v_{\nu}/L$, triggered by bosonic annihilation operators $d_{\nu,n_{q}}$.\\

\noindent Electrons are described by the electron field operator $\psi_{s}(x)$ expressed as~\cite{fabrizio}
\begin{equation}
\psi_s(x)=e^{ik_Fx}\psi_{s,+}(x)+e^{-ik_Fx}\psi_{s,-}(x)\, , \label{eq:optot}
\end{equation}
in terms of $\psi_{s,r}(x)$, the field operators for right ($r=+$) and left ($r=-$) moving electrons. Here $k_{F}=\pi N/2L$ is the system Fermi momentum. The operators $\psi_{s,\alpha}(x)$ satisfy the condition $\psi_{s,+}(x)=-\psi_{s,-}(-x)$ thus ensuring that $\psi_{s}(x)$ obeys {\em open boundary conditions} $\psi_{s}(0)=\psi_{s}(L)=0$. Within the bosonization language we have
\begin{equation}
\psi_{s,+}(x)=\frac{\eta_s}{\sqrt{2\pi\alpha}}e^{-i\theta_{s}}
\,e^{i\frac{\pi \Delta N_sx}{L}}e^{i\frac{\Phi_{\rho}(x)+s\Phi_\sigma(x)}{\sqrt{2}}}\,. \label{eq:opright}
\end{equation}
with $\Delta N_s=N_s-N/2$, $[\theta_{s}, N_{s'}]=i\delta_{s,s'}$ and $\alpha$ a cutoff length of the order of the inverse Fermi wavevector $\alpha\sim k_F^{-1}$. The operators $\eta_{s}$ satisfy $\eta_s\eta_{s'}+\eta_{s'}\eta_s=2\delta_{s,s'}$ ensuring the correct
anticommutation relations among fields with different spin. Finally the
boson fields $\Phi_{\rho}(x)$, $\Phi_{\sigma}(x)$ are
\begin{equation}
\Phi_{\nu}(x)\!=\!\sum_{n_{q}>0}\frac{e^{-\alpha q/2}}{\sqrt{g_{\nu}n_q}}
\left\{\left[\cos{(qx)}-ig_{\nu}\sin{(qx)}\right]d^\dag_{\nu,n_{q}}+\mathrm{h.c.}\right\}\, .\nonumber
\end{equation}
\subsection{Electron density}
\label{sec:dens}
Finite size and interaction effects induce peculiar spatial oscillations of the electron density
\begin{equation}
	\rho(x)=\sum_{s}\psi^{\dagger}_{s}(x)\psi_{s}(x)\, .\label{eq:rho1}
\end{equation}
The most important contributions are a long-wave (L) modulation, slowly varying on the scale of the dot, the Friedel contribution (F) oscillating with a period $\approx 2L/N$ and the Wigner one (W) oscillating at $L/N$, yielding
\begin{equation}
\rho(x)=\sum_{\chi=L,F,W}k_{\chi}\rho^{(\chi)}(x)\,,
\end{equation}
with
\begin{eqnarray}
\rho^{(L)}(x)&=&\sum_{r,s}\psi^{\dagger}_{s,r}(x)\psi_{s,r}(x)\,,\\
\rho^{(F)}(x)&=&\sum_{s}\left[e^{-2ik_{F}x}\psi^{\dagger}_{s,+}(x)\psi_{s,-}(x)+ \mathrm{h.c.}\right]\,,\\
\frac{\rho^{(W)}(x)}{\pi\alpha}&=&e^{-4ik_Fx}\psi^\dag_{+,+}\!(x)\psi_{+,-}\!(x)\psi^\dag_{-,+}\!(x)\psi_{-,-}\!(x)+\mathrm{h.c.}\, ,
\end{eqnarray}
and $k_{L}=1$, $k_{F}=1-\lambda$, $k_{W}=\lambda$.
\noindent Here, $\lambda\in[0,1]$ is an interaction-dependent model parameter which modulates the relative strength of Friedel and Wigner contributions to the density, and ensures the right boundary conditions $\rho(0)=\rho(L)=0$. Numerical investigations~\cite{bortz} suggest that for strongly interacting or diluted systems $\lambda\to 1$ while for weak interactions $\lambda\approx 0$.\\
\noindent In bosonized form one has~\cite{noiNJP13}
\begin{eqnarray}
\rho^{(L)}(x)&=&\!\!\frac{2k_F}{\pi}+\frac{\Delta N_{\rho}}{L}-\frac{\sqrt{2}}{\pi}\partial_x \varphi_\rho(x)-\frac{2g^2}{\pi}\partial_{x}h(x)\,,\\
\rho^{(F)}(x)&=&\!\!\!-\sum_{s}\frac{1}{2\pi}\partial_{x}\sin\left[\mathcal{L}_{s}(x)-2\varphi_{s}(x)\right]\, ,\\
\rho^{(W)}(x)&=&-\frac{1}{2\pi}\partial_{x}\sin\left[2\mathcal{L}\left(x\right)-2\sqrt{2}\varphi_\rho(x)\right]\,.
\end{eqnarray}
where
\begin{eqnarray}
\varphi_{\rho/\sigma}(x)&=&\frac{1}{2}\left[\Phi_{\rho/\sigma}(-x)-\Phi_{\rho/\sigma}(x)\right].\label{eq:varphi}\\
\varphi_s(x)&=&\frac{\varphi_\rho(x)+s\varphi_\sigma(x)}{\sqrt{2}},\\
\mathcal{L}(x)&=&2k_{F} x -2g^{2}h(x)\, ,\\
h(x)&=&\frac{1}{2}\tan^{-1}\left[\frac{\sin(2\pi x/L)}{e^{\pi\alpha/L}-\cos (2\pi x/L)}\right]\, ,\label{eq:effe}
\end{eqnarray}
and $\mathcal{L}_{s}(x)=\mathcal{L}(x)+2\pi\Delta N_{s}x/L$.\\
\noindent To proceed, we perform a thermal average over the dot degrees of freedom, which we assume thermalized and described by an equilibrium density matrix $\mathcal{D}=Z^{-1}\exp(-\beta H_{d})$, with $Z=\mathrm{Tr}\{\exp(-\beta H_{d})\}$ the corresponding partition function and $\beta^{-1}=k_{B}T$. We therefore introduce $\bar{\rho}(x)=\mathrm{Tr}\left\{\mathcal{D}\rho(x)\right\}$ and obtain~\cite{noiNJP13}
\begin{equation}
\bar{\rho}(x)=\sum_{\chi=L,F,W}k_{\chi}\left\langle\bar{\rho}^{(\chi)}(x)\right\rangle\,,
\end{equation}
with $\bar{\rho}^{(F)}(x)=\sum_{s}\bar{\rho}_{s}^{(F)}(x)$ and
\begin{eqnarray}
\bar{\rho}^{(L)}(x)&=&\frac{2k_F}{\pi}-\frac{2g^2}{\pi}\partial_{x}h(x)\, ,\\
\bar{\rho}^{(F)}_{s}(x)&=&-\frac{1}{2\pi}\partial_{x}\frac{\sin\left[\mathcal{L}_{s}\left(x\right)\right]}{\alpha_{\rho}(x)\alpha_{\sigma}(x)},\label{eq:rs}\\
\bar{\rho}^{(W)}(x)&=&-\frac{1}{2\pi}\partial_{x}\frac{\sin\left[2\mathcal{L}\left(x\right)\right]}{\alpha_{\rho}^{4}(x)}\,.\label{eq:rw}
\end{eqnarray}
Here, the bracket
\begin{equation}
\!\!\!\!\!\!\!\!\langle\mathcal{O}\rangle=\frac{1}{Z_{0}}\sum_{\Delta N_{\sigma}}\langle \Delta N_{\sigma}|e^{-\beta \frac{E_{\sigma}}{2}\Delta N_{\sigma}^{2}}\mathcal{O}|\Delta N_{\sigma}\rangle\,;\,	 Z_{0}=\sum_{\Delta N_{\sigma}}e^{-\beta\frac{E_{\sigma}}{2}\Delta N_{\sigma}^{2}}
\end{equation}
denotes the average over the dot spin zero modes. Furthermore, $\alpha_{\nu}(x)=\beta_{\nu}(x,x)$ with
\begin{equation}
\!\!\!\!\!\!\!\!\!\!\!\!\beta_{\nu}(x,y)=\exp\left[g_{\nu}\sum_{n>0}\frac{e^{-\frac{n\pi a}{L}}}{n}\sin\left(\frac{n\pi x}{L}\right)\sin\left(\frac{n\pi y}{L}\right)\coth\left(n\frac{T_{\nu}}{2T}\right)\right]\,,
\end{equation}
where $k_{B}T_{\nu}=\varepsilon_{\nu}$. One finds $\beta_{\nu}(x,y)=\prod_{p}\beta^{(p)}_{\nu}(x,y)$ with $p\in\mathbb{Z}$
\begin{equation}
\beta_{\nu}^{(p)}(x,y)=\left\{\sqrt{\frac{\sin^2\left[\frac{\pi(x+y)}{L}\right]+\sinh^2\left(\frac{\pi \alpha}{L}+|p|\frac{T_{\nu}}{T}\right)}{\sin^2\left[\frac{\pi(x-y)}{L}\right]+\sinh^2\left(\frac{\pi \alpha}{L}+|p|\frac{T_{\nu}}{T}\right)}}\right\}^{g_{\nu}}.
\end{equation}
The terms $\beta^{(0)}_{\nu}(x,y)$ represents the zero-temperature result, while terms with $|p|\geq 1$ are thermal corrections only relevant if $T\gtrsim |p|T_{\nu}$. Due to the constraint $k_{B}T\ll D_{\sigma}\ll D_{\rho}$, one has $\beta_{\rho}(x,y)\approx\beta_{\rho}^{(0)}(x,y)$ with an excellent precision.\\
\noindent The Friedel component of $\bar{\rho}(x)$ is composed by a superposition of terms oscillating at all wavelengths $2L/(N+\Delta N_{s})$, by virtue of the thermal average over the spin zero modes. For very small temperatures $T\ll T_{\sigma}$ the term with $\Delta N_{s}=0$ is the only relevant contribution to the average, leading to the standard Friedel oscillations with wavelength $2L/N$. For higher temperatures, partial cancellations due to the overlap of oscillating terms with different wavelengths occur, leading to a suppression of $\bar{\rho}^{(F)}(x)$. The Wigner term, on the other hand, is characterized by the wavelength $L/N$ and is robust in the temperature range explored in this paper.
\subsection{Pair correlation function}
\label{sec:pcf}
A tool to investigate in more details the correlations that develop into the quantum dot and its internal electronic structure is the PCF~\cite{vignale,2Dn11}
\begin{equation}
\!\!\!\!\!\!\!\bar{G}(x,x')=\frac{1}{\bar{\rho}(x)\bar{\rho}(x')}\sum_{s,s'}\mathrm{Tr}\left\{\mathcal{D}\psi_{s}^{\dagger}(x)\psi_{s'}^{\dagger}(x')\psi_{s'}(x')\psi_{s}(x)\right\}\, ,\label{eq:g2}
\end{equation}
which estimates the degree of correlation between the density at positions $x$ and $x'$ in the system. Due to the Pauli exclusion principle, $\bar{G}(x,x)<1$ while in general positive correlations can develop for $x'\neq x$. For an {\em uncorrelated} system (and $x\neq x'$), the trace in Eq.~(\ref{eq:g2}) decouples and $\bar{G}(x,x')=1$. Since the quantum dot is not translationally invariant, $\bar{G}(x,x')$ depends in a non-trivial way on both $x$ and $x'$. The PCF can be expressed in terms of the two-point density-density correlator $\mathrm{Tr}\left\{\rho(x)\rho(x')\right\}$. Expressed in terms of products involving the long-wave, Friedel and Wigner components of the density, the PCF becomes
\begin{equation}
\!\!\!\!\!\!\!\!\!\!\!\!\bar{G}(x,x')=\frac{1}{\bar{\rho}(x)\bar{\rho}(x')}\!\!\sum_{\chi,\chi'=L,F,W}\!\!k_{\chi}k_{\chi'}\left\langle \bar{G}^{(\chi,\chi')}(x,x')\right\rangle-\frac{C(x,x')}{\bar{\rho}(x)}
\end{equation}
with
\begin{eqnarray}
\!\!\!\!\!\!\!\bar{G}^{(L,L)}(x,x')&=&\bar{\rho}^{(L)}(x)\bar{\rho}^{(L)}(x')+\frac{2g\partial_{x,x'}^{2}}{\pi^2}\log\left[\beta_{\sigma}(x,x')\right]\\
\!\!\!\!\!\!\!\bar{G}^{(F,F)}_{s,s'}(x,x')&=&\!\!\!\!\!-\sum_{p=\pm}\frac{\partial_{x,x'}^{2}}{16\pi^2}\frac{p\cos\left[\mathcal{L}_{s}(x)+p\mathcal{L}_{s'}(x')\right]}{A(x,x')\beta_{\rho}^{2p}(x,x')\beta_{\sigma}^{2pss'}(x,x')}\label{eq:gff}\\
\!\!\!\!\!\!\!\bar{G}^{(W,W)}(x,x')&=&-\sum_{p=\pm}\frac{\partial_{x,x'}^{2}}{4\pi^2}\frac{p\cos\left[2\mathcal{L}(x)+2p\mathcal{L}(x')\right]}{\alpha_{\rho}^{4}(x)\alpha_{\rho}^{4}(x')\beta_{\rho}^{8p}(x,x')}\\
\!\!\!\!\!\!\!\bar{G}^{(F,L)}_{s}(x,x')&=&-\frac{\partial_{x,x'}^{2}}{\pi^2}\frac{\log\left[\beta_{\rho}(x,x')\right]\cos\left[\mathcal{L}_{s}(x')\right]}{\alpha_{\sigma}(x')\alpha_{\rho}(x')}\label{eq:glf}\\
\!\!\!\!\!\!\!\bar{G}^{(W,L)}(x,x')&=&-\frac{2\partial_{x,x'}^{2}}{\pi^2}\frac{\log\left[\beta_{\rho}(x,x')\right]\cos\left[2\mathcal{L}(x')\right]}{\alpha_{\rho}^{4}(x')}\\
\!\!\!\!\!\!\!\bar{G}^{(F,W)}_{s}(x,x')&=&-\sum_{p=\pm}\frac{\partial_{x,x'}^{2}}{8\pi^2}\frac{p\cos\left[\mathcal{L}_{s}(x)+2p\mathcal{L}(x')\right]}{\alpha_{\sigma}(x)\alpha_{\rho}(x)\alpha_{\rho}^{4}(x')\beta_{\rho}^{4p}(x,x')}\, .\label{eq:gfw}
\end{eqnarray}
Here we have introduced $A(x,x')=\alpha_{\rho}(x)\alpha_{\rho}(x')\alpha_{\sigma}(x)\alpha_{\sigma}(x')$ and
\begin{equation}
C(x,x')=\frac{L}{\pi}\frac{a}{x^2+a^2}\, .
\end{equation}
Furthermore, $\bar{G}^{(F,F)}(x,x')=\sum_{s,s'}\bar{G}^{(F,F)}_{s,s'}(x,x')$, $\bar{G}^{(L,F)}(x,x')=\sum_{s}\bar{G}^{(L,F)}_{s}(x,x')$, $\bar{G}^{(F,W)}(x,x')=\sum_{s}\bar{G}^{(F,W)}_{s}(x,x')$. Note finally that $\bar{G}^{(\chi',\chi)}(x,x')=\bar{G}^{(\chi,\chi')}(x',x)$.\\
\noindent Both the electron density $\bar{\rho}(x)$ and the PCF $\bar{G}(x,x')$ can be conveniently evaluated by means of standard computer algebra systems even for large $N$, which constitutes a great advantage of the analytical approach allowed by the Luttinger liquid formalism.
\section{Results}
\label{sec:res}
We begin by briefly recalling the behaviour of the electron density $\bar{\rho}(x)$ as a function of the interaction strength $g$ and of the temperature.
\begin{figure}[htbp]
\begin{center}
\includegraphics[width=8cm,keepaspectratio]{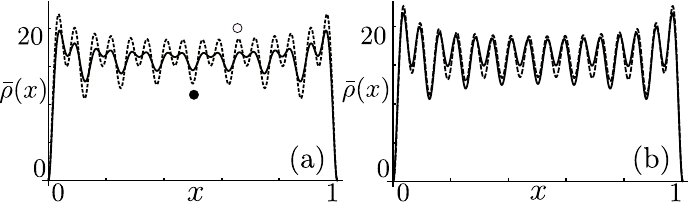}
\caption{Plot of $\bar{\rho}(x)$ (units $L^{-1}$) as a function of $x$ (units $L$) for $N=16$ and: (a) $k_{B}T=0.1 E_{\sigma}$ and $g=0.4$ (solid), $g=0.2$ (dashed); (b) $g=0.2$ and $k_{B}T=0.5 E_{\sigma}$ (solid), $k_{B}T=4 E_{\sigma}$ (dashed). In all panels, $\alpha=L/13\pi$ and $\lambda=0.5$. The dots in panel (a) denote the posititions chosen as one of the coordinates of the PCFs displayed in Fig.~\ref{fig:fig2}.}
\label{fig:fig1}
\end{center}
\end{figure}
Figure~\ref{fig:fig1}(a) shows $\bar{\rho}(x)$ for $N=16$ electrons in the dot at low temperature. For intermediate interactions, $g=0.4$, it exhibits $N/2$ peaks, corresponding to oscillations with the Friedel wavelength $2L/N$. For stronger interactions $g=0.2$, however, $N$ distinct peaks emerge signalling the incipience of Wigner correlations in the density with a wavelength $L/N$. The crossover between Friedel and Wigner is due to the different power-law scalings of these contributions, induced by the terms $\alpha_{\rho}^{-1}(x)\alpha_{\sigma}^{-1}(x)$ in Eq.~(\ref{eq:rs}) and $\alpha_{\rho}^{-4}(x)$ in Eq.~(\ref{eq:rw}). In particular, the Friedel oscillations roughly decay as $\sim(\alpha/L)^{(1+g)/2}$ while the Wigner term scales as $\sim(\alpha/L)^{2g}$: as a consequence, for $g\to 0$ Wigner oscillations tend to acquire their full amplitude while Friedel ones still remain suppressed. The superposition of Friedel and Wigner terms is however still visible in the density: the $N$ peaks are grouped in pairs, reminescent of the Friedel oscillation super-period $2L/N$. The situation is however different at higher temperature, as shown in Fig.~\ref{fig:fig1}(b). Indeed, as temperature is increased above $E_{\sigma}/k_{B}$, spin zero-mode excitations become thermally activated. This leads to the superposition of Friedel terms with different wavelengths already mentioned in the previous section. As a result, the Friedel contribution to $\bar{\rho}(x)$ gets damped in comparison to the Wigner one which is stable since $k_{B}T\ll E_{\rho}$. As a result, the peak grouping tends to disappear and Wigner correlations in the density tend to emerge even more as the comparison between the solid (low temperature) and the dashed (high temperature) lines in Fig.~\ref{fig:fig1}(b) shows.\\

\noindent Let us now turn to the discussion of the PCF.
\begin{figure}[htbp]
\begin{center}
\includegraphics[width=8cm,keepaspectratio]{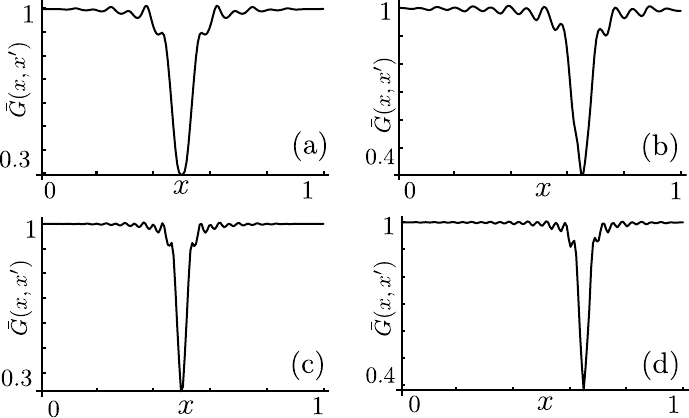}
\caption{Plot of $\bar{G}(x,x')$ as a function of $x$ (units $L$) for: (a) $N=16$ and $x'=0.5 L$ (solid dot in Fig.~\ref{fig:fig1}(a)); (b) $N=16$ and $x'=0.65 L$ (open dot in Fig.~\ref{fig:fig1}(a)); (c) same as in (a) but $N=30$; (d) same as in (b) but $N=30$. In all panels, $g=0.2$, $k_{B}T=0.5 E_{\sigma}$, $\lambda=0.5$, $v_{\sigma}/v_{\rho}=1/20$ and $\alpha=L/13\pi$ (a,b) or $\alpha=L/25\pi$ (c,d).}
\label{fig:fig2}
\end{center}
\end{figure}
Figure~\ref{fig:fig2} shows $\bar{G}(x,x')$ as a function of $x$ for different values of $x'$ and different particle numbers. In both panels (a) and (c) $x'$ sits on a minimum of $\bar{\rho}(x)$, denoted by the full dot in Fig.~\ref{fig:fig1}(a), while in panels (b) and (d) $x'$ is chosen on a maximum of $\bar{\rho}(x)$, the open dot in Fig.~\ref{fig:fig1}(a). All the four panels of Fig.~\ref{fig:fig2} display a set of common features. Most prominent is the presence of a {\em correlation hole} for $x\approx x'$, where $\bar{G}(x,x')\ll 1$. This negative correlation is due to the Pauli exclusion principle and the Coulomb repulsion among electrons. Away from the correlation hole, $\bar{G}(x,x')$ exhibits an oscillatory behavior alternating between positive and negative correlations of the position $x$ with the position $x'$. These oscillations are the hallmark of Wigner correlations within the system: indeed the number of oscillations is precisely $N-1$, reflecting the correlation of the electron at $x'$ with the other $N-1$ electrons in the molecule. Correlations disappear near the dot borders: indeed, we find that $\bar{G}(0,x')\approx \bar{G}(L,x')\approx 1$. As expected, Wigner correlations are more evident for lower particle numbers as the comparison between Figs.~\ref{fig:fig2}(a),(b) and Figs.~\ref{fig:fig2}(c),(d) confirms. However, significant Wigner correlations are present even for rather large numbers of particles $N\sim 30$. The correlations are stronger when $x'$ sits on a maximum of the single particle electron density and thus, from now on, we will always choose this coordinate accordingly.

\begin{figure}[htbp]
\begin{center}
\includegraphics[width=8cm,keepaspectratio]{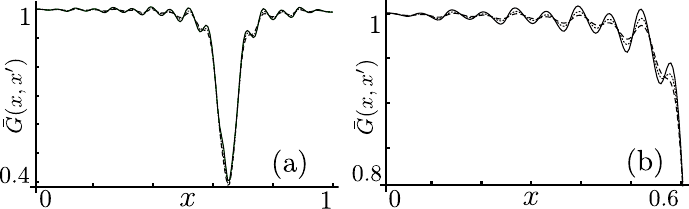}
\caption{(a) Plot of $\bar{G}(x,x')$ as a function of $x$ (units $L$) for $N=16$, $x'=0.65$ and $g=0.4$ (dashed), $g=0.3$ (dotted) and $g=0.2$ (solid). (b) Zoom of the plot in (a) to the left of the correlation hole, for $0\leq x\leq 0.6$. Other parameters: $\lambda=0.5$, $k_{B}T=0.5 E_{\sigma}$, $v_{\sigma}/v_{\rho}=1/20$ and $\alpha=L/13\pi$.}
\label{fig:fig3}
\end{center}
\end{figure}
Figure~\ref{fig:fig3}(a) displays the behaviour of the PCF as the interaction strength is increased. While the correlation hole is essentially dominated by the Pauli principle and is almost unaffected by the variations in $g$, the spatial oscillations of $\bar{G}(x,x')$ are enhanced as $g$ is decreased in analogy to the behaviour of $\bar{\rho}(x)$. This effect is particularly evident in Fig.~\ref{fig:fig3}(b) which shows a zoom of the PCF to the left of the correlation hole: the amplitude of oscillations is increased. Also, a tendency towards the grouping of the oscillations can be detected, reminescent of the behaviour of the density at low temperatures.

\begin{figure}[htbp]
\begin{center}
\includegraphics[width=8cm,keepaspectratio]{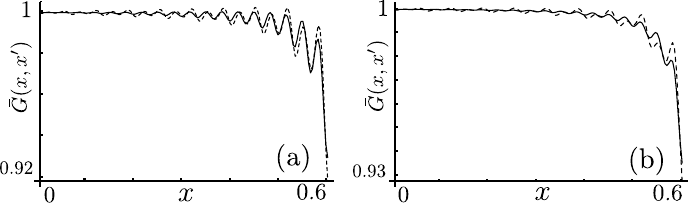}
\caption{Plot of $\bar{G}(x,x')$ as a function of $x$ (units $L$) for $x'=0.65$ to the left of the correlation hole $0\leq x\leq 0.6$ for $k_{B}T=0.5 E_{\sigma}$ (dashed) or $k_{B}T=20 E_{\sigma}$ (solid) and: (a) $g=0.2$; (b) $g=0.4$. In all panels, $\lambda=0.5$, $v_{\sigma}/v_{\rho}=1/20$ and $\alpha=L/25\pi$.}
\label{fig:fig4}
\end{center}
\end{figure}
To investigate this issue, we report in Fig.~\ref{fig:fig4} results for the PCF in the case of low temperature ($k_{B}T=0.5 E_{\sigma}$, dashed line) and high temperature ($k_{B}T=20 E_{\sigma}$, solid line) for the case of $N=30$. Both for strong interactions, see Fig.~\ref{fig:fig4}(a), and for moderate ones, see Fig.~\ref{fig:fig4}(b), the tendency towards a grouping of the peaks of $\bar{G}(x,x')$ disappears as temperature is increased above $E_{\sigma}/k_{B}$. This confirms what already suggested by the behaviour of the electron density, namely that Friedel oscillations are less robust than Wigner correlations and that an intermediate temperature regime exists in which finite-size effects are suppressed leading to an emergence of Wigner correlations. The vanishing of Friedel effects can be ultimately traced back to the superposition of several different terms oscillating with different wavelengths in Eqns.~(\ref{eq:gff}),(\ref{eq:glf}), and (\ref{eq:gfw}). We want to stress that the high temperature regime analyzed here is still within the validity range $k_{B}T<D_{\sigma}$.
\section{Conclusions}
\label{sec:conc}
In this paper we have evaluated the PCF of a strongly interacting one-dimensional quantum dot. Employing the spin-coherent Luttinger liquid model within the bosonization language, analytical expressions have been obtained, which can be easily evaluated even for very large numbers of electrons. By investigating its behaviour as a function of the interaction strength and of the system temperature, we were able to confirm that at low temperature Wigner correlations get enhanced over finite-size Friedel oscillation, even though the latter still modulate the spatial behaviour of the PCF. Increasing the temperature above the spin zero-mode excitation gap leads to a superposition of oscillatory patterns in the terms of the pair correlation function which involve the Friedel oscillations, leading to their suppression in comparison to the Wigner ones. An intermediate temperature regime is then reached in which Wigner effects emerge over finite-size ones.\\

\noindent {\em Acknowledgements.} Financial support by the EU-FP7 via ITN-2008-234970 NANOCTM is gratefully acknowledged.

\end{document}